\documentclass[amsmath,amssymb,amsfonts,superscriptaddress,twocolumn]{revtex4}

\usepackage{graphicx,color,float}

\newcommand{\rb}{\overline{r}}
\newcommand{\s}{\overline{s}}
\newcommand{\ub}{\overline{u}}
\newcommand{\rhob}{\overline{\rho}}
\newcommand{\ab}{a}
\newcommand{\xs}{x_\mathrm{s}}

\newcommand{\pp}[2]{\frac{\partial #1}{\partial #2}}

\begin{document}

\title{Hydrodynamic Optical Soliton Tunneling}

\author{P.~\surname{Sprenger}}
\email{patrick.sprenger@colorado.edu}
\author{M. A.~\surname{Hoefer}}
\email{hoefer@colorado.edu}
\affiliation{Department of Applied Mathematics, University of Colorado,
  Boulder CO 80309, USA}
\author{G. A.~\surname{El}}
\email{g.el@lboro.ac.uk}
\affiliation{Department of Mathematical Sciences, Loughborough University,
  Loughborough, LE11 3TU, UK}
\date{\today}

\begin{abstract}
  A conceptually new notion of hydrodynamic optical soliton tunneling
  is introduced in which a dark soliton is incident upon an evolving,
  broad potential barrier that arises from an appropriate variation of
  the input signal.  The barriers considered include smooth
  rarefaction waves and highly oscillatory dispersive shock
  waves. Both the soliton and the barrier satisfy the same
  one-dimensional defocusing nonlinear Schr\"odinger (NLS) equation,
  which admits a convenient dispersive hydrodynamic interpretation.
  Under the scale separation assumption of nonlinear wave (Whitham)
  modulation theory, the highly nontrivial nonlinear interaction
  between the soliton and the evolving hydrodynamic barrier is
  described in terms of self-similar, simple wave solutions to an
  asymptotic reduction of the Whitham-NLS partial differential
  equations. One of the Riemann invariants of the reduced modulation
  system determines the characteristics of a soliton interacting with
  a mean flow that results in soliton tunneling or trapping. Another
  Riemann invariant yields the tunneled soliton's phase shift due to
  hydrodynamic interaction. Under certain conditions, soliton
  interaction with hydrodynamic barriers gives rise to new effects
  that include reversal of the soliton propagation direction and
  spontaneous soliton cavitation, which further suggest possible
  methods of dark soliton control in optical fibers.
\end{abstract}

\pacs{
  05.45.Yv, 
  }

\maketitle

\section{Introduction}

The tunneling of wavepackets incident upon a potential barrier is a
defining quantum mechanical property \cite{landau_quantum_1965}. The
linear phenomenon can be extended to nonlinear solitonic wavepackets
or solitons--localized, unchanging waveforms in which nonlinear and
dispersive effects are in balance. In the original consideration of a
soliton incident upon a potential barrier, it was found that the
soliton can losslessly pass, or tunnel, through a localized repulsive
or attractive potential \cite{newell_nonlinear_1978}. The theoretical
connection of this so-called soliton tunneling with quantum mechanical
tunneling was established in an optical setting in
\cite{serkin_soliton_1993} where a bright optical pulse propagating in
an optical fiber with anomolous dispersion was transmitted through a
localized defective region of normal dispersion--the analog of a
potential barrier. Soliton tunneling has been studied theoretically in
some detail in recent years in various physical systems including
optical media
\cite{anderson_tunneling_1994,zhong_soliton_2010,wang_nonlinear_2008,yang_spatial_2008},
nematic liquid crystals
\cite{assanto_optical_2009,peccianti_escaping_2008} and matter waves
in Bose-Einstein condensates (BEC)
\cite{rodas-verde_controllable_2005,loomba_controlling_2014}. Recent
experiments observed the nonlinear analogs of some linear quantum
features including nonlinear scattering \cite{linzon_nonlinear_2007},
reflection and ejection \cite{barak_observation_2008} and soliton
tunneling \cite{marest_longitudinal_2017}.

In the focusing (anomalous dispersion) regime, nonlinear optical plane
wave propagation is subject to modulational instability with respect
to long wavelength perturbations \cite{zakharov_modulation_2009}.  In
contrast, plane wave propagation in the normal dispersion (defocusing)
regime is stable and, remarkably, exhibits many features
characteristic of fluid motion \cite{carusotto_quantum_2013}. The
dispersive effects in such a ``fluid of light'' are due to diffractive
or chromatic properties of the medium.  The dispersive hydrodynamic
behavior of light propagation has been considered and observed in a
number of works, see, e.g., \cite{wan_dispersive_2007,
  garnier_incoherent_2013, xu_dispersive_2017}.

Robust features of the diffraction of laser light in a nonlinear,
defocusing medium and matter waves in a repulsive BEC include dark
solitons, moving depression waves whose width is proportional to the
coherence length $l$ of the medium.  In addition to solitons, these
media also support spatially extended, smooth configurations that can
exhibit wavebreaking and the spontaneous emergence of highly
oscillatory dispersive shock waves (DSWs) \cite{el_dispersive_2016}.
Optical DSWs have been observed in both bulk media
\cite{wan_dispersive_2007} and optical fibers
\cite{conti_observation_2009}.  While the DSW oscillatory length scale
is also the medium's microscopic coherence length $l$, DSWs exhibit
expanding, rank-ordered oscillations spanning a larger, macroscopic
coherence length scale $L$, which increases with time.  This latter
length scale also characterizes non-oscillatory hydrodynamic flows
such as expansion or rarefaction waves (RWs) and compressive Riemann
waves that have recently been observed in optical fibers in the
context of wavebreaking control \cite{wetzel_experimental_2016}.  The
scale separation $l \ll L$, a natural characterization of dispersive
hydrodynamics \cite{biondini_dispersive_2016}, enables a mathematical
description of DSWs via nonlinear wave, Whitham averaging
\cite{whitham_linear_1974,el_dispersive_2016}, while RWs are described
by the long-wave, dispersionless limit of the original equations.

Despite the fact that solitons, RWs and DSWs are well known,
fundamental features of dispersive media, soliton-RW and soliton-DSW
interactions, have been mostly overlooked. As we show, these
interactions motivate an alternative notion of optical tunneling
whereby a dark soliton incident upon a spatially extended hydrodynamic
barrier in the form of a DSW or a RW can penetrate through to the
other side of the evolving hydrodynamic structure.  Thus, in contrast
to the traditional notion of soliton tunneling through an externally
imposed barrier, hydrodynamic soliton tunneling corresponds to the
full penetration and emergence of the soliton through an intrinsic
hydrodynamic state that evolves {\it according to the same equation as
  the soliton}.  This generalizes the understanding of a soliton as a
coherent, particle-like entity that can interact elastically with
other solitons \cite{zabusky_interaction_1965} and dispersive
radiation \cite{ablowitz_note_1982} to one that can also interact with
nonlinear hydrodynamic states and emerge intact, i.e., without
fissioning or radiation, albeit with a different amplitude that
results from a change in the background mean flow.

In this paper, we analyze the tunneling of solitons through
hydrodynamic states within the framework of the integrable, defocusing
nonlinear Schr\"{o}dinger (NLS) equation, which is an accurate model
for nonlinear light propagation in single mode optical fibers with
normal dispersion \cite{boyd_nonlinear_2013}.  We invoke the scale
separation $l \ll L$ inherent to Whitham modulation theory in order to
derive a system of asymptotic equations that describe the interaction
between narrow dark solitons and evolving, broad hydrodynamic
barriers. We obtain the conditions on the incident soliton amplitude
and hydrodynamic mean flow density and velocity for tunneling. One of
the fundamental properties of hydrodynamic soliton tunneling is {\it
  hydrodynamic reciprocity} whereby the tunneling through RWs and DSWs
is described by the same set of conditions in spite of the very
different interaction dynamics. This general property of {\it
  solitonic hydrodynamics} has been recently formulated and
experimentally confirmed for a fluid system
\cite{maiden_solitonic_2017}.  We also show that tunneling is not
always possible and that the soliton can be absorbed or trapped within
the hydrodynamic flow. Moreover, we find that soliton interaction with
hydrodynamic states can lead to reversal of the soliton's propagation
direction and spontaneous soliton cavitation.

Our analysis can be applied to a large class of dispersive
hydrodynamic systems, including dispersive Eulerian equations
\cite{hoefer_shock_2014,el_dispersive_2016} which have broad
applications. 
The particular case of optical hydrodynamic soliton tunneling
considered here could be observed, for example, within the
experimental setting described in \cite{xu_dispersive_2017} for the
generation of DSWs and RWs in optical fibers.

\section{Problem formulation}
We consider the defocusing NLS equation 
\begin{equation}
  \label{eq:0}
  i \psi_t = - \frac{1}{2} \psi_{xx} + |\psi|^2 \psi ,
\end{equation}
where in the context of fiber optic propagation, $t$ is the
longitudinal coordinate in the fiber, $x$ is the retarded time, and
$\psi(x,t)$ is the complex-valued, slowly-varying envelope of the
electric field. All variables are nondimensionalized to their typical
values. 
See, e.g., \cite{xu_dispersive_2017} for a detailed description of NLS
normalizations and typical values of physical parameters pertinent to
the regimes considered here.

Equation \eqref{eq:0} can be written in dispersive hydrodynamic form
via the transformation $\psi = \sqrt{\rho} e^{i \phi}$, $u = \phi_x$
\begin{equation}
  \label{eq:1}
  \rho_t + (\rho u)_x = 0, \quad
  u_t + uu_x + \rho_x = \left ( \frac{\rho_{xx}}{4\rho} -
    \frac{\rho_x^2}{8 \rho^2} \right )_x ,
\end{equation}
where $\rho$ is the optical power and $u$ is the chirp.  In terms of
the hydrodynamic interpretation of these quantities, we will refer to
$\rho$ as a mass density and $u$ as a flow velocity (see, e.g.,
\cite{el_dispersive_2016}). Within this setting, the normalized
coherence length is $l = \rho_0^{-1/2}$ where $\rho_0$ is a typical
density scale.  The coherence length is an intrinsic scale that, along
with the coherence time $\tau = \rho_0^{-1}$, corresponds to a scaling
invariance of the hydrodynamic equations \eqref{eq:1}. In BECs, $l$ is
known as the healing length \cite{pitaevskii_bose-einstein_2003}.

Equation \eqref{eq:1} admits the localized, dark soliton solution 
\begin{equation}
  \label{eq:3}
  \begin{split}
    \rho(x,t) &= \rhob - a \,\mathrm{sech}^2[\sqrt{a}(x-c t -x_0)],\\
    u(x,t) &= \ub \pm \sqrt{\rhob -a}[1-\rhob/\rho(x,t)],
  \end{split}
\end{equation}
where $a$ is the maximum deviation from the mean density $\rhob$,
$\ub$ is the mean flow velocity, and $c = \ub \pm \sqrt{\rhob - a}$ is
the soliton amplitude-speed relation.  The $\pm$ in \eqref{eq:3} is
due to the bi-directional nature of the NLS equation as a dispersive
hydrodynamic system \eqref{eq:1}. When $a = \rhob$, the soliton is
called a black soliton because its minimum is a zero density,
cavitation point.

The typical tunneling problem consists of a soliton incident on a
fixed potential barrier, either due to a change in the medium or an
external effect.  However, the spatio-temporal barriers considered
here evolve according to the same equation that describes the dynamics
of the medium. For an optical fiber with homogeneous, normal
dispersion, this corresponds to a time-dependent input signal that
results in both a soliton and a large-scale barrier.  We assume that
the hydrodynamic mean flow $(\rhob,\ub)$ that develops from the
initial data varies on much longer length and time scales $L \gg l$,
$T \gg \tau$, respectively.  In this regime, the third order
dispersive term in \eqref{eq:1} is negligible, resulting in the
long-wave, dispersionless, quasilinear equations for the mean flow $\rho \to \rhob$, $u \to \ub$
\begin{equation}
  \label{eq:4}
  \rb_t + \frac{1}{2}(3\rb + \s) \rb_x = 0, \quad \s_t +
  \frac{1}{2}(\rb + 3\s) \s_x = 0,
\end{equation}
written in diagonal form where
\begin{equation}
  \label{eq:2}
  \rb = \ub/2-\sqrt{\rhob}, \quad \s = \ub/2 + \sqrt{\rhob},
\end{equation}
are the Riemann invariants.  In fact,
Eqs.~\eqref{eq:4} are the shallow water equations in one dimension and
RWs are determined exclusively by the constancy of $\rb$ or $\s$
\cite{leveque_finite_2002}. Remarkably, the same constant Riemann
invariant determines the loci of simple wave DSWs
\cite{el_dispersive_2016}, in contrast to viscous shock waves of
classical fluid dynamics, whose loci are determined by the
Rankine-Hugoniot conditions \cite{leveque_finite_2002}.

We consider the problem of a dark soliton \eqref{eq:3} incident upon a
barrier that evolves from step initial data in the mean flow
$\rhob(x,0)$, $\ub(x,0)$, where
\begin{align} \label{eq:5} \rhob(x,0) = \begin{cases}\rhob_- \quad x <
    0 \\ \rhob_+ \quad x > 0\end{cases}\hspace{-0.4cm},\quad \ub(x,0)
  = \begin{cases}\ub_- \quad x < 0 \\ \ub_+ \quad x >
    0\end{cases}\hspace{-0.4cm} .
\end{align}
As we will show, the long-time evolution of soliton-hydrodynamic
barrier interaction is determined by the far-field flow conditions
$\rhob_\pm$ and $\ub_\pm$.  Therefore, our theory generalizes to
soliton tunneling through arbitrary hydrodynamic barriers with given
far-field conditions.

\begin{figure}[t!]
  \centering
  \includegraphics{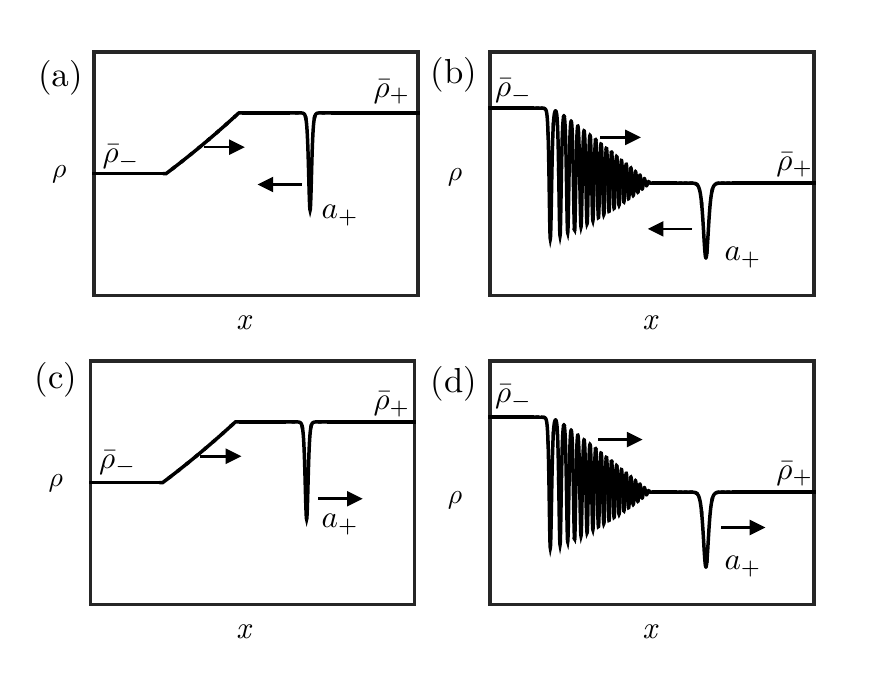}
  \caption{Hydrodynamic soliton tunneling configurations. a) RW
    overtaking soliton b) DSW overtaking soliton c) RW soliton
    collision d) DSW soliton collision. }
  \label{fig:configurations}
\end{figure}

The step initial conditions \eqref{eq:5} generally evolve into a
combination of two waves: RWs and/or DSWs each characterized by a
simple-wave locus of the dispersionless limit system \eqref{eq:4}
\cite{el_decay_1995,el_dispersive_2016}.  Therefore, we shall be
imposing a simple-wave constraint on the initial mean flow data
\eqref{eq:5}, i.e., we assume that either $\rb(x,0)$ or $\s(x,0)$
found from (\ref{eq:2}) is constant across $x=0$ so that the mean flow
will evolve into a single expanding hydrodynamic wave, either a RW or
a DSW. Due to the bi-directional nature of the NLS equation, there are
four distinct configurations, defined by the direction of the jump (up
or down) of the Riemann invariant $\rb$ or $\s$ across $x=0$. We will
focus on the two cases that result in a RW or a DSW when $\rb(x,0)$ is
constant.  These two configurations along with an incident dark
soliton moving to the left or right define four basic cases of
hydrodynamic soliton tunneling considered here and shown in
Fig.~\ref{fig:configurations}.

We pause briefly to note some common terminology in the nonlinear
waves literature \cite{el_dispersive_2016}.  The RW and DSW depicted
in Fig.~\ref{fig:configurations} are referred to as a 2-RW and a
2-DSW, respectively because their characteristic wave speeds
degenerate to the \textit{fastest} long wave speed
$\ub_0 + \sqrt{\rhob_0}$ when $\rhob_+, \rhob_- \to \rhob_0$,
$\ub_+, \ub_- \to \ub_0$.  The other two cases where $\s(x,0)$ is
constant correspond to a 1-RW or a 1-DSW because their speeds
degenerate to the slowest long wave speed $\ub_0 - \sqrt{\rhob_0}$.
These 1-waves can be obtained from the 2-waves considered here with
the reflection invariance $x \to -x$, $u \to -u$ of Eqs.~(\ref{eq:1}).

To describe how the mean flow couples to the soliton amplitude during
the interaction, we utilize Whitham modulation theory
\cite{whitham_linear_1974}. The general framework for Whitham
modulation theory encompasses slow modulation on the space and time
scales $L$ and $T$ of a periodic wave's parameters, which lead to a
system of quasi-linear partial differential equations (PDE) for the
parameter evolution. For the NLS equation, the modulation equations
are a system of four equations that can be written in diagonal form
\cite{pavlov_hamiltonian_1993,forest_geometry_1986,kamchatnov_nonlinear_2000,el_dispersive_2016}
\begin{equation}\label{eq:riemann}
  \pp{r_i}{t} + V_i({\bf{r}})\pp{r_i}{x} = 0 , \quad i = 1, \dots , 4.
\end{equation}
The Riemann invariants $\mathbf{r}$ satisfy
$r_4 \ge r_3 \ge r_2 \ge r_1$ and vary on the much larger
spatiotemporal scales $L$ and $T$ than the scales $l$ and $\tau$ of
the soliton \eqref{eq:3}.  The characteristic velocities are computed
via
\begin{equation}\label{eq:charvel}
  V_i({\bf{r}}) = \left( 1- \frac{\lambda}{\partial_i \lambda}\partial_i\right)U,
\end{equation}
where $\partial_i = \pp{}{r_i}$, and
\begin{equation}\label{LV}
  \lambda = \frac{2 K(m)}{\sqrt{(r_4 - r_2)(r_3 - r_1)}}, \quad U=
  \frac{1}{2}\sum\limits_{j = 1}^{4} r_j, 
\end{equation}
are the wavelength and phase velocity, respectively. Here, $K(m)$ is
the complete elliptic integral of the first kind and
$m = \left[(r_2-r_1)(r_4 - r_3)]/[(r_4-r_2)(r_3-r_1)\right].$ The
characteristic velocities exhibit the ordering $V_i \le V_j$ if
$1\le i \le j \le 4$. The wave amplitude is $a = (r_2-
r_1)(r_4-r_3)$. By setting all but one Riemann invariant constant, we
obtain an equation for a simple wave of modulation, which we call a
$j$-wave, where $j$ is the index of the non-constant, varying Riemann
invariant.

The equations \eqref{eq:riemann} are consistent with the wave
conservation law
\begin{equation}\label{eq:wc}
  k_t +(kU)_x=0, \quad k= 2\pi/\lambda.
\end{equation}

Soliton-mean field interaction is described by the soliton limit of
the NLS-Whitham Eqs.~\eqref{eq:riemann}, which is achieved when
$r_2=r_3$ (see, e.g., \cite{el_dispersive_2016}).  By analyzing the
expression Eq.~\eqref{eq:charvel} for the characteristic velocities in
the soliton limit $r_2=r_3$, it is possible to establish that the
limiting modulation system consists of shallow water equations
\eqref{eq:5} where $\s=r_4$, $\rb = r_1$ and the equation for the
merged Riemann invariant $r_3$ is \cite{el_dispersive_2016}
\begin{equation}\label{eq:r3}
  r_{3,t} + \frac 12 (\rb + 2r_3+ \s)r_{3,x} = 0, 
\end{equation}
with 
\begin{equation}\label{eq:r3_definition}
  r_3 = \bar{u}/2 \pm \sqrt{\bar{\rho} - a},
\end{equation}
where the two signs are due to bi-directionality (cf., the second
formula in Eq.~\eqref{eq:3}).

Thus, effectively, Eq.~\eqref{eq:r3} is the equation for the soliton
amplitude $a(x,t)$. Crucially for our consideration, the soliton
amplitude here is a spatiotemporal {\it field}, satisfying a PDE,
while in standard soliton perturbation theories
\cite{kivshar_dynamics_1989}, the soliton amplitude has only a
temporal dependence that satisfies an ODE along the soliton
trajectory. The trajectory and dynamics of a single soliton from the
amplitude field can be interpreted as the introduction of a fictitious
train of non-interacting solitons of the same amplitude and some small
wavenumber $0 < k \ll 1$, which necessarily satisfies the wave
conservation equation \eqref{eq:wc} with
$U = \frac 12 \left(\rb + 2r_3 + \s\right) = c$, the soliton
amplitude-speed relation.  Using the limiting system (\ref{eq:4}),
\eqref{eq:r3}, the wave conservation equation \eqref{eq:wc} can be
written in diagonal, Riemann invariant form
\begin{align}
  \label{eq:wcRiemann}
  \begin{split}
    (kp)_t + c \left(kp\right)_x = 0,\\
    p = \displaystyle \exp \left(-
      \int\limits_{\s_0}^{\s}\frac{\frac{d c_\mathrm{s}}{d
          \s}}{\frac{1}{2} \left(\rb + 3 s \right) - c} d s\right),
  \end{split}
\end{align}
where $\s_0$ is some fixed reference value, e.g., $\s_-$.

Thus, the initial conditions \eqref{eq:5} for the hydrodynamic barrier
should be complemented by similar conditions for the soliton amplitude
field and the small wavenumber,
\begin{align}
  \label{eq:ICs}
  a(x,0) = \begin{cases} a_- \quad x < 0 \\  a_+ \quad x >
    0\end{cases} \hspace{-0.4cm},\quad  k(x,0) = \begin{cases} k_-
    \quad x < 0  \\ k_+ \quad x > 0\end{cases}\hspace{-0.4cm},
\end{align}
where only the incident amplitude $a_+$ is given at the onset (recall
the configurations in Fig.~\ref{fig:configurations}).

The hydrodynamic soliton tunneling problem then consists in finding
(i) the transmitted soliton amplitude $a_-$; (ii) the stretching
(contraction) coefficient $k_+/k_-$ for the soliton train that
determines the soliton phase shift due to tunneling.

Concluding this section, we note that the long wave limit of the Whitham equations demonstrates that while the soliton amplitude is coupled to the evolving mean flow, the mean flow itself evolves \emph{independently} of additional localized nonlinear waves.

\section{Hydrodynamic soliton tunneling}

We shall consider the basic tunneling configurations depicted in
Fig.~\ref{fig:configurations}, which are defined by constancy of one
of the hydrodynamic Riemann invariants $\rb, \s$ in the step initial
data \eqref{eq:5}.  Without loss of generality, one can choose
$(\rhob_-,\ub_-) = (1,0)$, the remaining configurations can be deduced
from scaling, Galilean shifts, and reflection symmetries associated
with Eq.~\eqref{eq:1}.

We note that, given step initial conditions, the hydrodynamic system
\eqref{eq:4} is valid only if the resulting wave is a RW. This implies
that the reduced single-phase modulation system \eqref{eq:4},
\eqref{eq:r3}, \eqref{eq:wcRiemann} describes only soliton-RW
interactions (cases (a) and (c) in
Fig.~\ref{fig:configurations}). Indeed, soliton DSW interaction is
more complicated and generally requires consideration of two-phase NLS
modulation equations \cite{forest_geometry_1986}. Remarkably, however,
we will show that the soliton-DSW tunneling conditions for cases (b)
and (d) can be found from the soliton-RW tunneling conditions via {\it
  hydrodynamic reciprocity}.

\begin{figure}
\includegraphics[scale=0.3]{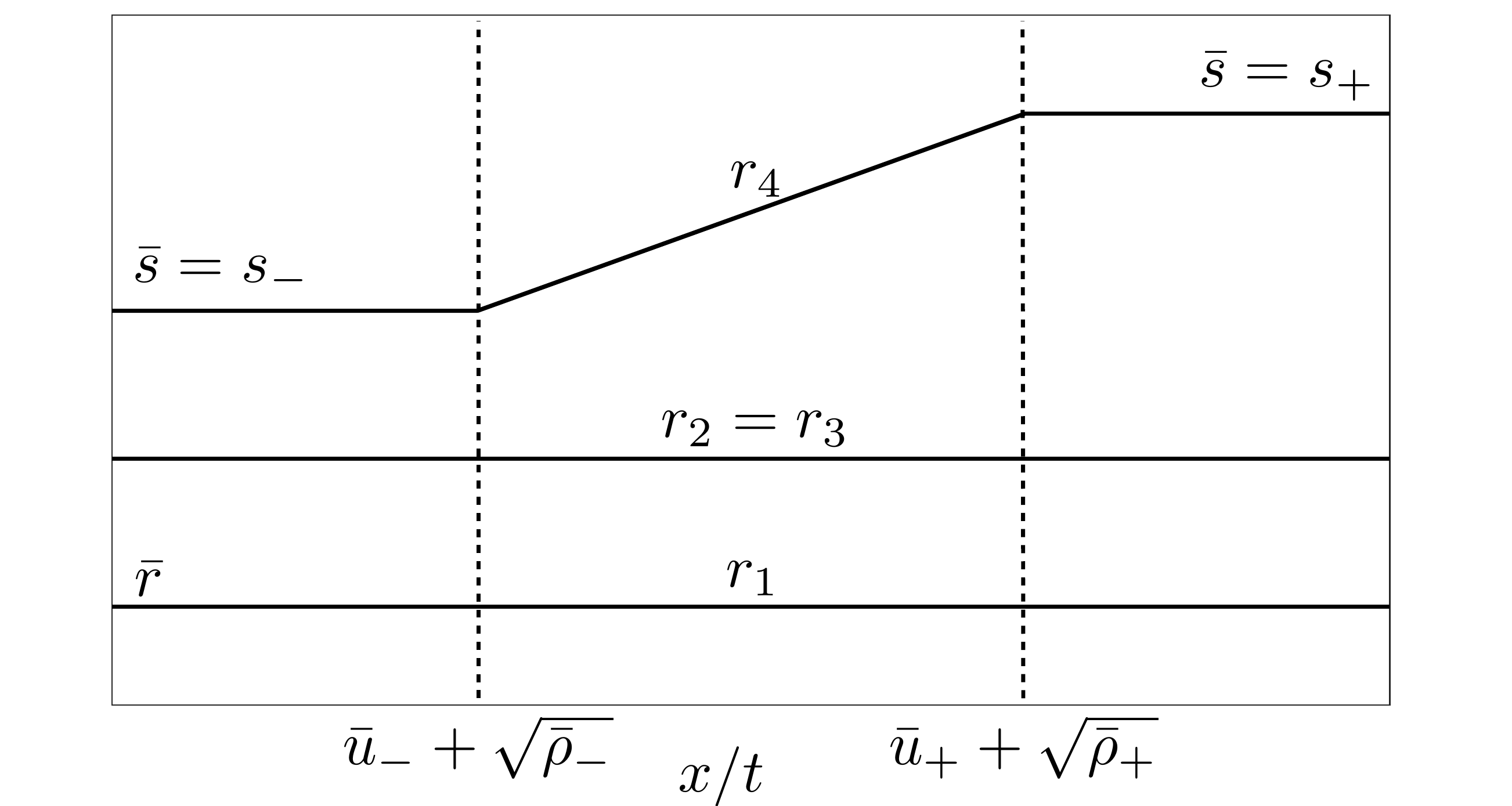}
\caption{Hydrodynamic soliton tunneling configuration of the Riemann
  invariants for soliton-RW interactions.  
    }
\label{fig:Riemann_invariants}
\end{figure}

Let us assume that $\rb(x,0)$ has no jump across $x=0$ and the jump in
$\s(x,0)$ resolves into a RW. The modulation equations (\ref{eq:4}),
\eqref{eq:r3}, \eqref{eq:wcRiemann} with step initial conditions for
$\s$, $r_3$ and $kp$ found from (\ref{eq:5}), \eqref{eq:ICs} imply the
simple wave solution in which $\rb=r_1$, $r_3$, and $kp$ are constant for all
$(x,t)$ but $\s = r_4$ is varying in a self-similar fashion,
$\s = \s(x/t)$.  This 4-wave modulation solution describes the
hydrodynamic tunneling configurations (a) and (c) in
Fig.~\ref{fig:configurations}.  An example 4-wave evolution is shown
in Fig.~\ref{fig:Riemann_invariants}.

The tunneling problem now essentially reduces to finding the constant values of $r_3$ and $kp$ given the constant value of $\rb = \ub_+/2 - \sqrt{\rhob_+}  =  \ub_-/2 - \sqrt{\rhob_-} = -1$ and the initial jump for $\s$ found from \eqref{eq:5} so that the Riemann invariants resemble those in Fig.~\ref{fig:Riemann_invariants}. The solution for $\s(x/t)$ will then define the soliton trajectory through a hydrodynamic RW barrier. 

The requirement of constancy of $r_3$ defined by
Eq.~\eqref{eq:r3_definition} when evaluated with (\ref{eq:5}) and
(\ref{eq:ICs}) yields a simple algebraic expression for the
transmitted soliton amplitude through a RW
\begin{equation}\label{eq:7}
  a_- = a_+ -2(\sqrt{\rho_+} \pm \sqrt{\rhob_+-\ab_+} )(\sqrt{\rho_+}
  - 1). 
\end{equation}
Importantly, tunneling through the hydrodynamic barrier requires
$0 < a_- \leq 1$. The $\pm$ in Eq.~\eqref{eq:7} corresponds to the two
branches of $r_3$ with ``$-$'' corresponding to the collision case
depicted in Fig.~\ref{fig:configurations}(a) and ``$+$'' the
overtaking case depicted in Fig.~\ref{fig:configurations}(c).  The
transmitted, or tunneled, soliton amplitude-speed relation is then
\begin{align}\label{eq:8}
  \begin{split}
    c_{-} &= \pm \sqrt{1-a_-}\\
    & = \frac{1}{2}\left(\rb + r_3 + \s_-\right),
  \end{split}
\end{align}
with $a_-$ given by Eq.~\eqref{eq:7}. The expression for the soliton
velocity $c_{-}$ in terms of Riemann invariants is a convenient
representation that inherently incorporates the appropriate sign
$\pm$.  We shall also explore implications of constancy of $kp$.


\begin{figure}[H]
\centering
\includegraphics[scale=0.25]{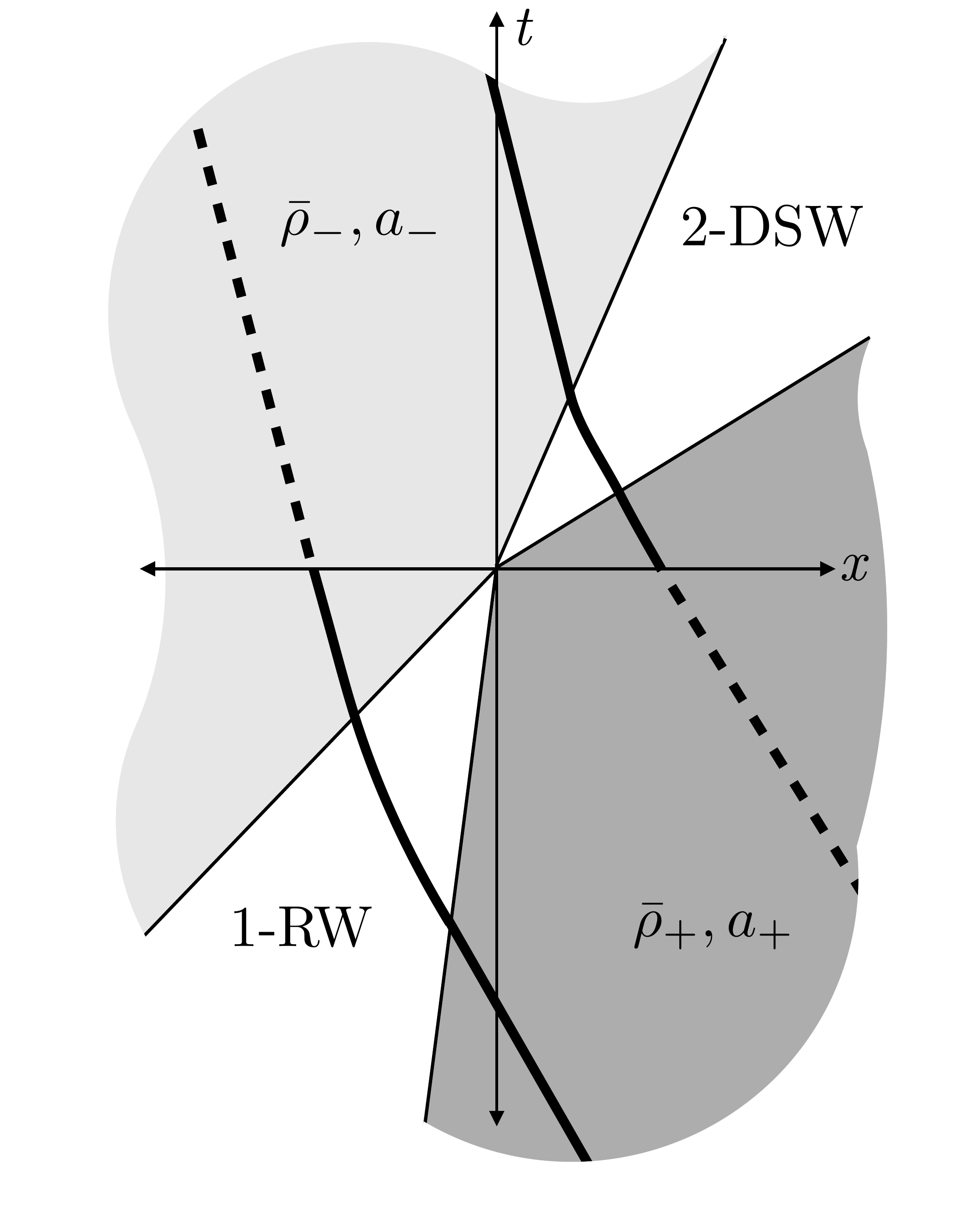}
\caption{Time reversibility of initial data $(\rhob_+,a_+)$ and
  $(\rhob_-,a_+)$ with $\rhob_+ < \rhob_-$. Forward temporal evolution
  results in soliton interaction with a 2-DSW (upper half plane) and
  backward evolution results in soliton interaction with a
  1-RW. Soliton trajectories are depicted with solid and dashed
  curves.  
   }
\label{fig:reciprocity_contour}
\end{figure}
The formulae \eqref{eq:7} and \eqref{eq:8}, in spite of their
simplicity, exhibit a number of remarkable implications.  These
include soliton tunneling, soliton trapping, the spontaneous emergence
of a cavitation point, and soliton direction reversal.  Furthermore,
the obtained conditions incorporate the fundamental notion of
hydrodynamic reciprocity established for uni-directional systems of
the Korteweg-de Vries (KdV) type in \cite{maiden_solitonic_2017}. This
states that the tunneling conditions {\it are the same for both the RW
  and DSW}.  This concept enables the application of Eqs.~(\ref{eq:7})
and (\ref{eq:8}) to soliton-DSW interaction.

To extend the reciprocity result of \cite{maiden_solitonic_2017} to
the hydrodynamic optical tunneling considered here, we consider a
general case where the left background state is $(\rhob_-, \ub_-)$
(not necessarily $(1,0)$) and take either $\rb(x,0)$ or $\s(x,0)$
constant.  This generalization will require consideration of both
branches of $r_3$ in Eq.~\eqref{eq:r3_definition} and in the tunneling
condition \eqref{eq:7}.  Hydrodynamic reciprocity ultimately results
from the time and space reversibility of the NLS equation
(\ref{eq:0}).

We first consider the soliton-DSW interaction case where
$\rhob_- > \rhob_+$ so that the DSW is known as a 2-DSW
\cite{el_dispersive_2016}.  The soliton is initially located to the
right of the DSW so that the hydrodynamic transition across the DSW
satisfies a 4-wave modulation curve in which $\rb = r_1 = const$ (see
\cite{el_decay_1995,el_dispersive_2016})
\begin{equation}
  \label{eq:6}
  \ub_-  - \ub_+ = 2(\sqrt{\rhob_-} - \sqrt{\rhob_+}).
\end{equation}
The nonlinear superposition of a soliton and a DSW can be achieved by
considering the modulation of two-phase (quasi-periodic) solutions of
the NLS equation (\ref{eq:0}) \cite{forest_geometry_1986}.  Therefore,
a description of the full soliton-DSW modulation would require the
integration of the two-phase Whitham equations.  However, we can
determine all the results of soliton-DSW interaction by invoking
continuity of the modulation solution for negative time.  

If we now consider $t \to -t$ for the Whitham modulation equations
(\ref{eq:riemann}), then the characteristic velocities $-V_i$ are
re-ordered.  The same initial data $\rhob_- > \rhob_+$ and the locus
(\ref{eq:6}) corresponds to the generation of a 1-RW.  If a soliton of
amplitude $a_-$ is initialized to the left of the RW, then soliton-RW
interaction is determined by the constancy of $r_3$ so that the
tunneled soliton amplitude satisfies
\begin{equation}
  \label{eq:1-wave}
  a_+ = \ab_- - 2 \left(\sqrt{\rhob_-} \pm \sqrt{\rhob_--\ab_-}\right)
  \left(\sqrt{\rhob_-}-\sqrt{\rhob_+}\right),
\end{equation}
where the $\pm$ corresponds to the same branch of $r_3$ that is taken.
The relation (\ref{eq:1-wave}) corresponds to a 1-wave modulation of
the time-reversed Whitham equations.  This is a global relationship
that must also hold for the corresponding 4-wave soliton-DSW
modulation of the non-reversed Whitham equations due to continuity of
the modulation solution away from the origin.  This analysis is
pictured in Fig.~\ref{fig:reciprocity_contour} where, for negative
time, a soliton-RW interaction is pictured and a soliton-DSW
interaction is shown for positive time.  

\begin{figure}
 \includegraphics{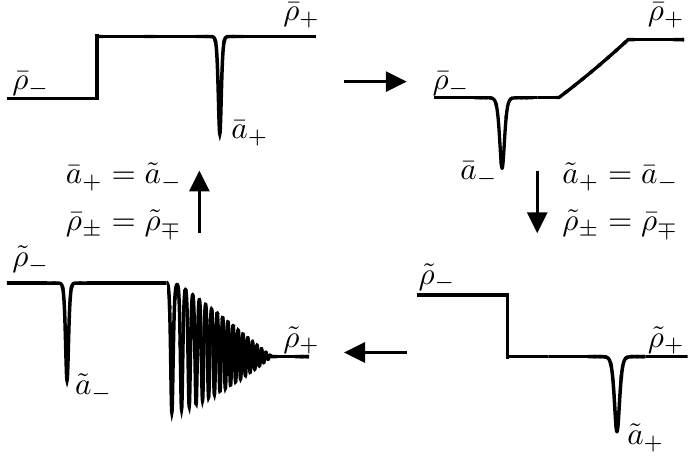}
 \caption{Sketch of configurations demonstrating hydrodynamic
   reciprocity. Horizontal arrows refer to temporal evolution and
   vertical arrows connote the transformation to the reciprocal
   initial condition.}
 \label{fig:reciprocity}
\end{figure}
Equation (\ref{eq:1-wave}) can be inverted to obtain $a_-$ in terms of
$a_+$ and $\rhob_\pm$.  If we set $\rhob_- = 1$ and $\ub_- = 0$, then
Eq.~(\ref{eq:1-wave}) and Eq.~(\ref{eq:7}) are equivalent.  The
tunneling condition (\ref{eq:7}) is the same for both soliton-RW and
soliton-DSW interaction.

Another way to understand hydrodynamic reciprocity is schematically
pictured in Fig.~\ref{fig:reciprocity}.  Rather than reversing time,
this Figure depicts spatial reversal.  A soliton of amplitude $a_+$
initially placed to the right of a jump with $\rhob_- < \rhob_+$
results in soliton interaction with a 2-RW and $a_-$ satisfying
Eq.~(\ref{eq:1-wave}).  Now, consider a spatially reversed jump with
$\tilde{\rho}_\pm = \rhob_\mp$ so that
$\tilde{\rho}_- > \tilde{\rho}_+$.  With a soliton of amplitude
$\tilde{a}_+ = a_-$ initially placed on the right, the soliton
interaction with a 2-DSW results in the tunneled amplitude
$\tilde{a}_- = a_+$.  This is the bi-directional generalization of the
uni-directional hydrodynamic reciprocity condition noted in
\cite{maiden_solitonic_2017}.


In what follows, we compare the modulation theory predictions for
hydrodynamic optical soliton tunneling with numerical simulations of
Eq.~\eqref{eq:0} for initial data comprised of a smoothed step
Eq.~\eqref{eq:4} and a soliton. We use a standard 6th order finite
difference spatial discretization with Dirichlet boundary
conditions. Time evolution is achieved with the standard 4th order
Runge-Kutta method. The numerical evolution was validated by the
numerical evolution of the exact solitary wave solution on a uniform
background Eq.~\eqref{eq:3}.

\begin{figure}
\includegraphics[scale=0.4]{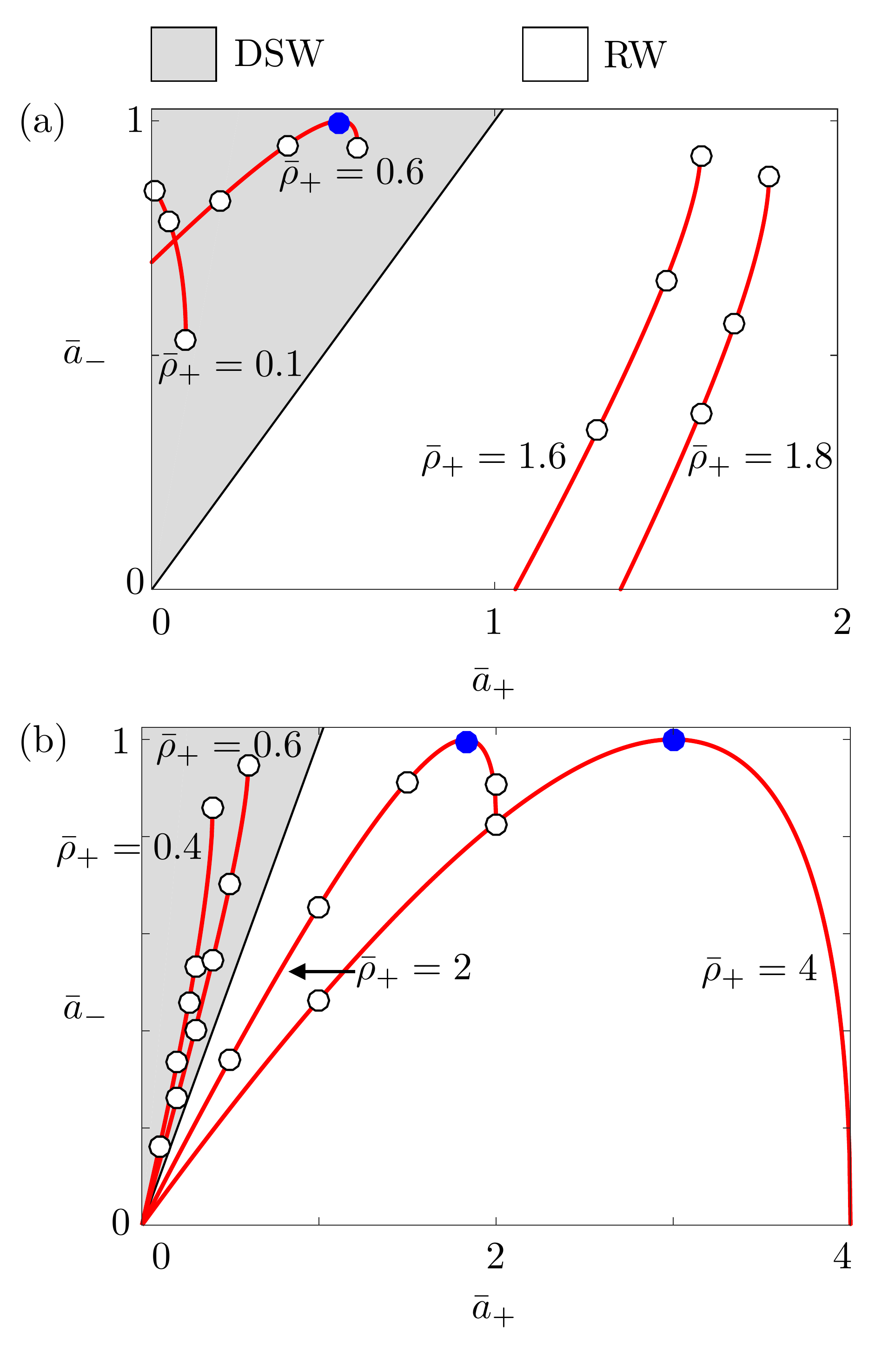}
\caption{Comparison between the tunneling relation (\ref{eq:7}) (solid
  curves) and direct numerical simulations of the NLS equation (dots)
  with smoothed, step initial data defined by $\rhob_+ > 0$ and a
  soliton of amplitude $a_+$. (a) The overtaking cases in
  Fig.~\ref{fig:configurations}(c,d). (b) The collision cases in
  Fig.~\ref{fig:configurations}(a,b).  Filled dots correspond to the
  emergence of a black soliton. The grey regions correspond to
  soliton-DSW tunneling and white regions correspond to soliton-RW
  tunneling.}
\label{fig:tunneling_amps}
\end{figure}

\begin{figure}[t]
\centering
\includegraphics[scale=0.10]{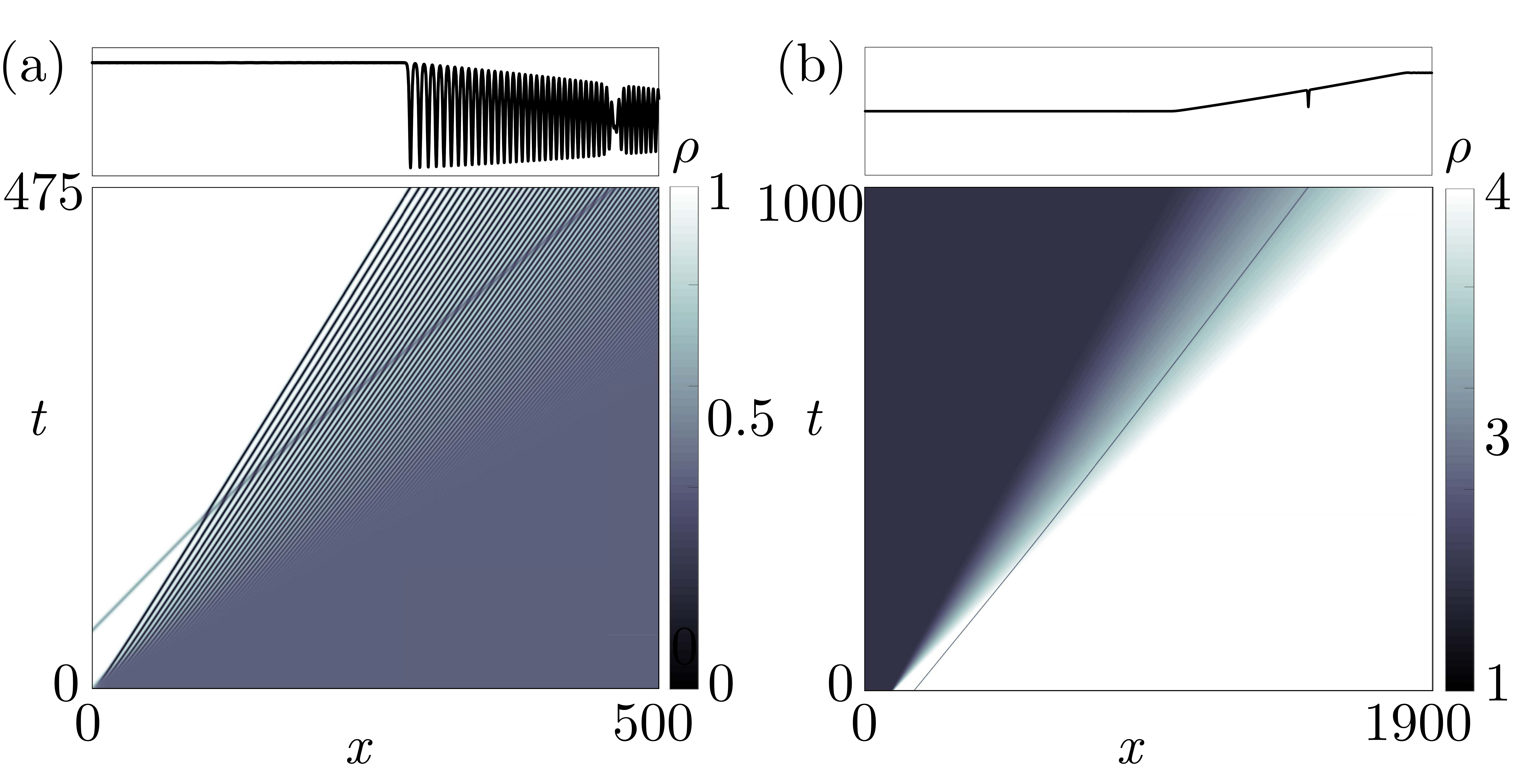}
\caption{Numerical simulation of hydrodynamically trapped
  solitons. (a) Soliton of amplitude $a_- = 0.25$ sent into a 2-DSW
  with $\rhob_+ = 0.5$. (b) Soliton of amplitude $a_+ = 1$ overtaken
  by RW with $\rho_+ = 4$. }
\label{fig:trapped_soliton}
\end{figure}

Comparisons between the transmitted soliton amplitude predicted by
Eq.~\eqref{eq:7} and numerical simulations are given in
Fig.~\ref{fig:tunneling_amps}, showing excellent agreement. When the
tunneling relation \eqref{eq:7} is not satisfied for $a_\pm>0$, the
soliton will become trapped within the spatially extended hydrodynamic
state.  Trapping then results in the soliton acting as a nonlinear
modulation of the hydrodynamic structure.  Examples of a soliton
trapped in a hydrodynamic barrier are shown in
Fig.~\ref{fig:trapped_soliton} where the soliton was unable to pass
through the RW or DSW for long simulation times. Soliton-DSW trapping
can be viewed as the formation of a ``defect'' in the locally periodic
DSW structure, analogous to the soliton defects of KdV cnoidal waves
considered in \cite{kuznetsov_stability_1975}.  In contrast to
classical optical soliton tunneling in which the localized pulse can
be reflected by a barrier with sufficient energy, this is not possible
in the context of hydrodynamic optical tunneling.

The simplest tunneling configuration is that of a solitary wave though
a RW because the evolution of the macroscopic structure of the RW can
be determined by standard methods applied to the modulation
Eqs.~(\ref{eq:4}) and (\ref{eq:r3}).  The RW evolution, in terms of
the Riemann invariants is
\begin{equation}
  \label{eq:RW_evol}
  \s_{\mathrm{RW}} (x,t) = \begin{cases} \s_- \quad &x < V_- t \\
    \frac 13 \left(2\frac{x}{t} - \rb\right)\quad & V_- t  \leq x \leq
    V_+t\\ 
    \s_+ & V_+t < x\end{cases},
\end{equation}
where $V(\s,\rb)= \frac{1}{2}(3 \s + \rb)$ and
$V_\pm = V(\s_\pm ,\rb)$ are the edge speeds of the centered RW
\cite{leveque_finite_2002}. We note that small amplitude linear
oscillations may be present at an edge of the RW in the full NLS
dynamics due to dispersive regularization but these oscillations decay
and have negligible influence on the asymptotic RW behavior.  
The trajectory of the soliton center $x_{\mathrm{s}}$ is a
characteristic (2- or 3-characteristic) of the Whitham modulation
equations \eqref{eq:riemann} that satisfies the initial value problem
\begin{equation}
  \label{eq:10}
  \frac{d \xs}{dt} = c\left(\rb,\s,r_3\right), \quad \xs(0) = x_+.
\end{equation}
Here, $x_+$ is the location of the solitary wave at $t= 0$ and the
location of the step (\ref{eq:5}) is taken to be $x = 0$; $c$ is the
soliton amplitude-speed relation \eqref{eq:8} written in terms of
Riemann invariants.  A direct integration of \eqref{eq:10} results in
the location of the solitary wave tunneling through a rarefaction wave
\begin{align}
  \label{eq:11}
  \xs(t) = \begin{cases}x_+ + c_+ t \quad  &x \leq t_1\\
    \displaystyle \frac{(\rb + r_3)t + 3 \left(\s_+ -  r_3\right)t_1^{2/3}t^{1/3}}{2}
    &t_1 < x  < t_2\\
    x_- + c_- t & t \geq t_2 
  \end{cases}, 
\end{align}
where
\begin{align*}
  t_1 &=x_+/(\s_+ - r_3),\\ t_2 &=(\s_+-r_3)^{3/2}(\s_- -
                                  r_3)^{-3/2}t_1, \\x_-
      &=(\s_+-r_3)^{1/2}(\s_- - r_3)^{-1/2}x_+, \\
  c_\pm &= \frac{1}{2}\left(\rb + 2r_3 + \s_\pm \right) .
\end{align*}
The effective phase shift of the soliton center through a RW is given
by the difference in the $x$-intercepts of the linear soliton
trajectories post and pre-hydrodynamic interaction.
\begin{equation}
  \label{eq:13}
  x_+ - x_- = \left(1- \sqrt{\frac{\s_+ - r_3}{\s_- - r_3}}\right)x_+.
\end{equation}
An alternative, instructive way to determine the interaction phase
shift is to analyze the additional modulation equation
\eqref{eq:wcRiemann} that describes the evolution of the wavenumber
$0 < k \ll 1$ in a train of well separated, non-interacting solitons
with the amplitude field $a(x,t)$. Given the mean flow
$\s_\mathrm{RW}(x,t)$ in \eqref{eq:RW_evol}, the amplitude field
$a(x,t)$ is determined by the constancy of $\rb$ and $r_3$ in
Eq.~\eqref{eq:r3_definition}.  The soliton phase shift now follows
from the requirement of constancy of the Riemann invariant $pk$ of
Eq.~\eqref{eq:wcRiemann} across the initial step \eqref{eq:5},
\eqref{eq:ICs}. Indeed, equating the values of $pk$ at both sides of
the initial step we find the ratio $k_+/k_- = x_-/x_+$, which
determines the stretching (contraction) of the soliton wavetrain at
leading order \cite{maiden_solitonic_2017},
\begin{equation}
  \label{eq:phaseShiftIntegral}
  \begin{split}
    \frac{k_+}{k_-} &= \frac{x_-}{x_+} = \exp \int\limits_{\s_-}^{\s_+}
    \frac{\displaystyle \frac{d c}{d \s}}{\frac{1}{2} \left(\rb +
        3\s\right) - c} d \s, \\
    &= \sqrt{\frac{\s_+ - r_3}{\s_- - r_3}},
  \end{split}
\end{equation}
where the first term in the denominator is the characteristic speed
associated with $\s$. This simpler approach yields the same result as
that obtained from Eq.~\eqref{eq:11}.

\begin{figure}[t]
  \centering
  \includegraphics[scale=0.10]{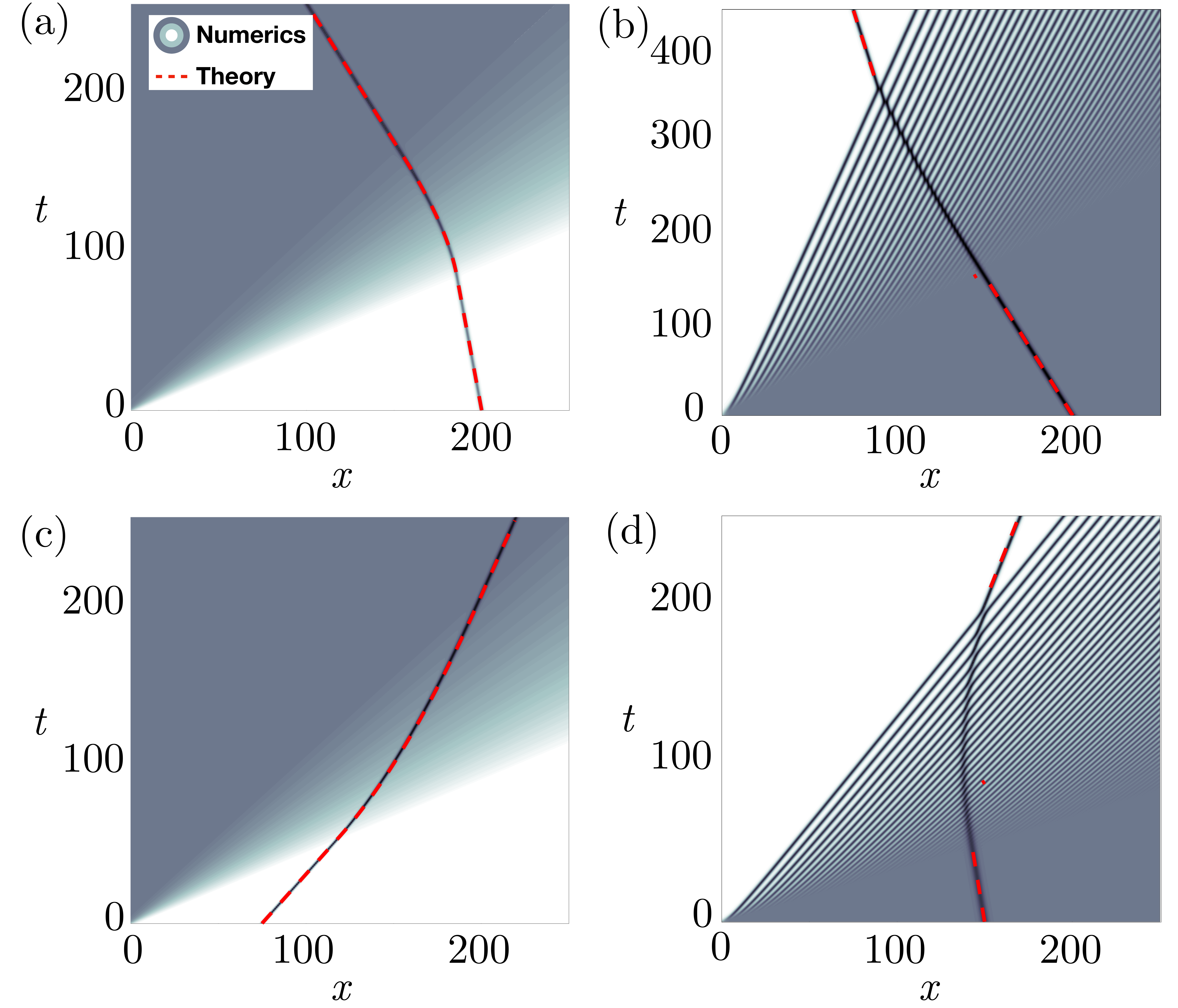}
  \caption{Tracking a soliton through a RW and DSW via equations
    \eqref{eq:11} and \eqref{eq:16}. (a) $\rhob_+$ = 2, $a_+ = 1$ and
    the sign of $r_3$ in Eq.~(\ref{eq:r3_definition}) is ``$-$''. (b)
    $\rhob_+$ = 0.5, $a_+ = 0.5$, the $r_3$ sign is ``$-$''.  (c)
    $\rhob_+ = 2$, $a_+ = 2$, the sign of $r_3$ is ``$+$''. (d)
    $\rhob_+$ = 0.5, $a_+ = 0.4$, the sign of $r_3$ is
    ``$+$''. Predictions from Eqs.~\eqref{eq:11} and \eqref{eq:16} are
    the red dashed curves, the contours are from direct numerical
    simulation of Eq.~(\ref{eq:0}).}
  \label{fig:soliton_tracking}
\end{figure}

We can now invoke the notion of hydrodynamic reciprocity-the
surprising fact that the interaction of the soliton with a RW is the
same as that with a DSW at the macroscopic level. In addition to the
tunneling relation (\ref{eq:7}), the phase shift \eqref{eq:13} also
applies to soliton-DSW interaction. The macroscopic properties of the
DSW itself--leading harmonic edge speed and trailing soliton edge
speed--are determined by an analysis of the single phase Whitham
equations in place of the direct integration that was possible in the
RW case. The distinguished edge speeds of the DSW are given by
\cite{gurevich_dissipationless_1987}
\begin{align}
  \label{eq:15}
  \begin{split}
    V_{-,\mathrm{DSW}} &= 2 \sqrt{\rhob_+}- 1, \quad \\ 
    V_{+,\mathrm{DSW}} & = \ub_+ \frac{\rhob_+-8
      \sqrt{\rhob_+}+8}{2-\sqrt{\rhob_+}}.  
  \end{split}
\end{align}
Incorporating the soliton phase shift
Eq.~(\ref{eq:phaseShiftIntegral}) results in the soliton trajectory
before and after interaction
\begin{align}
  \label{eq:16}
  x_{s,\mathrm{DSW}} = \begin{cases}x_+ + \frac{1}{2}\left(\rb + 2r_3 +
      \s_+\right)  t\quad &x\leq t_1\\
    x_- + \frac{1}{2}\left(\rb + 2r_3 + \s_-\right) t & x \geq t_{2}.  
  \end{cases},
\end{align}
where now, $t_1$, $t_2$ are determined by equating the pre and
post-interaction soliton trajectories with the appropriate DSW edge
velocities from Eq.~\eqref{eq:15}. Comparisons with numerical
simulations of soliton-DSW interactions are shown in
Figs.~\ref{fig:soliton_tracking}(b,d) with excellent agreement. The
trajectory prediction Eq.~\eqref{eq:16} also correctly captures the
phenomenon of soliton direction reversal shown in
Fig.~\ref{fig:soliton_tracking}(d).

The transition to a different mean flow across the hydrodynamic
barrier not only results in a controllable soliton trajectory but also
the generation of transmitted solitons of pre-specified amplitudes
(cf., Eq.~\eqref{eq:7}).  For specific initial configurations of the
tunneling problem, we predict and numerically observe the spontaneous
development of a black soliton that exhibits cavitation or a null in
the density at the soliton minimum. Black soliton solutions are
characterized in the normalization considered here by an amplitude
$a_- = 1$ with an associated $\pi$ phase jump across the soliton
minimum. In the reference frame chosen, the soliton velocity on the
left flow is given by $c_{-} = 0$.  The phenomenon of so-called self
cavitation of dispersive shock waves was theoretically predicted in
\cite{el_decay_1995} and both a zero density point and the associated
$\pi$ phase jump was observed experimentally for the dam break problem
of spin waves in a defocusing magnetic material
\cite{janantha_observation_2016}. Zero density points were also
observed in an optical ``photon fluid'' \cite{xu_dispersive_2017}. The
interaction of a dark soliton with a mean flow then gives a
fundamentally new mechanism for generating a cavitation point in the
flow.

\begin{figure}[H]
  \centering
  \includegraphics[scale=0.10]{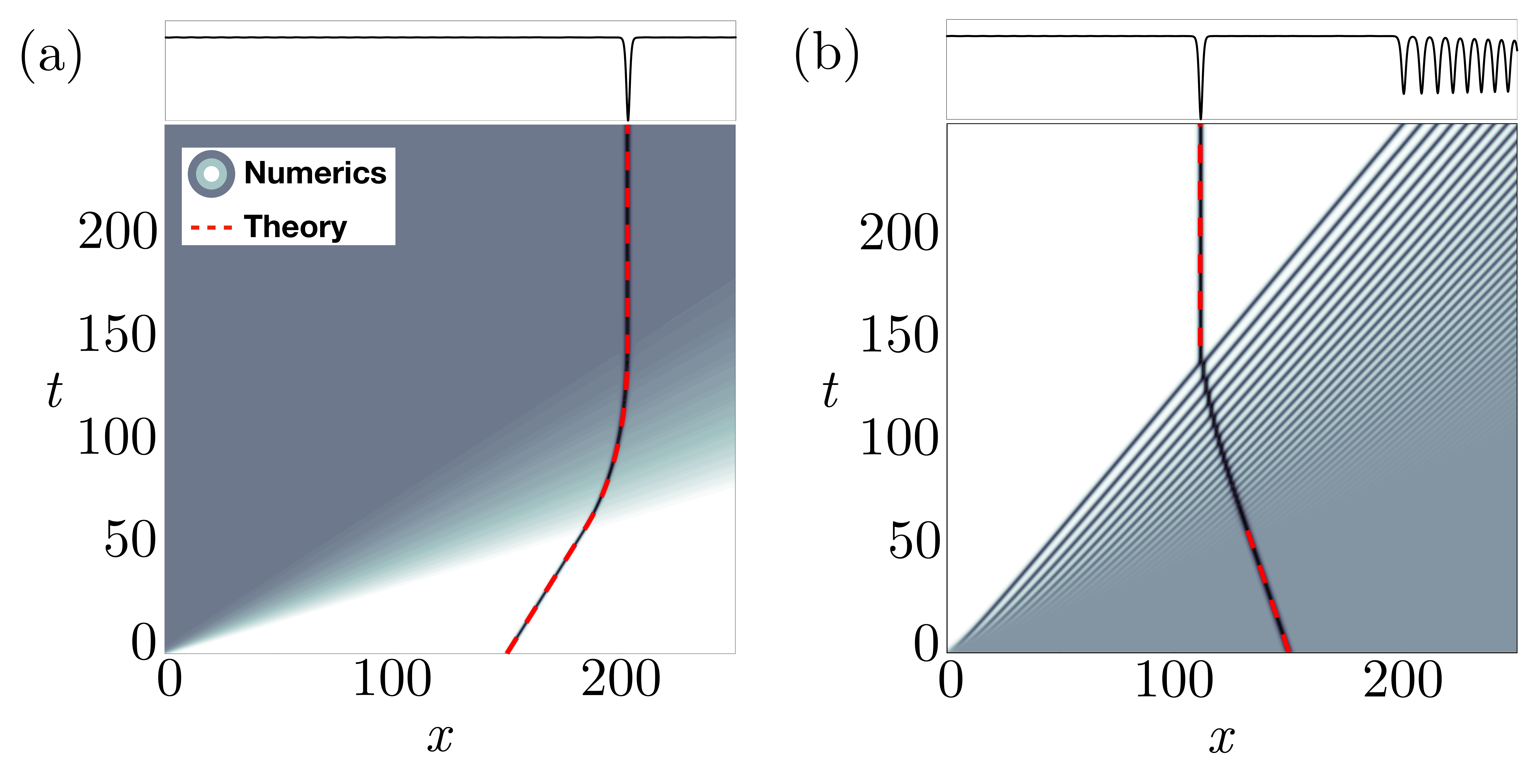}
  \caption{Examples of the emergence of a black soliton after
    tunneling in the characteristic plane. The initial configurations
    are (a) RW collision case with $\rhob_+ = 2$, (b) DSW overtaking
    case with $\rhob_+ = 0.6$.  Initial soliton amplitudes are chosen
    so that $a_- = 1$ in \eqref{eq:7}. Numerically computed soliton
    trajectories (contours) are compared against theoretical
    predictions of Eqs.~(\ref{eq:11}) and \eqref{eq:16}. The snapshots
    of the intensity $\rho$ at $t = 225$ are shown above the contour
    plots.}
\end{figure}

\section{Conclusion}

In this work, we have introduced a new notion of hydrodynamic optical
soliton tunneling where a localized, depression wave or dark soliton
is incident on a spatio-temporal hydrodynamic barrier.  Under the
assumptions of nonlinear wave, Whitham modulation theory, the
evolution of the inhomogeneous mean flow decouples from the soliton so
that, at the leading order macroscopic level, the flow is wholly
unaltered by the presence of the local pulse. The solution is found to
be a self-similar simple wave of a system of quasilinear partial
differential equations whose characteristics determine both the mean
flow and the soliton trajectory. The self-similar simple wave obtained
evolves from an initial step in the flow to either a single DSW or a
RW but the approach generalizes to any initial state that limits to
different constants as $x \to \pm \infty$, which define soliton
tunneling conditions.

The main result of this work is encompassed in the tunneling and phase
relations given by Eqs.~\eqref{eq:7} and
(\ref{eq:phaseShiftIntegral}). They determine the transmitted soliton
amplitude, speed, and position in terms of only the incident soliton
amplitude, its position, and the hydrodynamic flow in the far-field.
The known soliton trajectory and amplitude following interaction
provide a mechanism for soliton control via interaction with a
spatially extended mean flow.

The notion of hydrodynamic reciprocity identified earlier in
\cite{maiden_solitonic_2017} for scalar, KdV type systems and
generalized here to the NLS case allows one to investigate a complex
soliton-DSW interaction by studying the simpler case of soliton-RW
interaction. Reciprocity implies that, although the tunneling of a
soliton through a DSW involves a complex interaction with rapid
nonlinear oscillations, they are unimportant for determining the
resulting amplitude, velocity and shift of the solitary wave post
interaction. The methodology presented here to track the trajectory of
the soliton only requires knowledge of the far field boundary
conditions and hence this approach can be extended to more general
initial configurations.We also note that the developed theory is not restricted to integrable NLS dynamics and can be generalized to more general cases of hydrodynamic optical soliton tunneling described by non-integrable versions of the defocusing NLS equation, e.g. with saturable nonlinearity, using the methods of \cite{el_resolution_2005,hoefer_shock_2014,el_dispersive_2016}

\begin{acknowledgments}
  This work was partially supported by NSF CAREER DMS-1255422 (M.A.H.)
  and EPSRC grant EP/R00515X/1 (G.A.E.).
\end{acknowledgments}

\bibliographystyle{unsrtnat}



\end{document}